# ¿CÓMO SE AFECTA LA CINÉTICA DE ANTICUERPOS MONOCLONALES POR LA VARIACIÓN DE PARÁMETROS FÍSICOS?

# HOW IS THE MONOCLONAL ANTIBODIES KINETIC AFFECTED BY CHANGES OF THEIR PHYSICAL PARAMETERS?

Camilo Delgado-Correal[1], Carolina Daza[2] and Heidy Alexandra Lizarazo-Pérez[3]

*1mcdelgadoc@unal.edu.co Centro Internacional de Física-CIF, Colombia*
[2] *ycdazac@unal.edu.co Departamento de Física, Universidad Nacional de Colombia*
[3] *halizarazop@unal.edu.co Facultad de Medicina, Universidad Nacional de Colombia*

**RESUMEN:** El estudio de los anticuerpos monoclonales (MAb) constituye un campo de gran interés para la ciencia médica, por ejemplo, agentes anti-TNF (infliximab y adalimumab) constituyen una importante herramienta para el manejo de trastornos autoinmunes e inflamatorios en la actualidad. Caracterizar los parámetros físicos que puedan estar involucrados en el comportamiento cinético de los anticuerpos monoclonales en su medio de acción (plasma sanguíneo) y el estudio de la interacción que éstos presentan con el antígeno, pueden ser de gran ayuda para aumentar la cinética de los MAb y mejorar su eficacia. En este trabajo nos centramos en la descripción física de la cinética de transporte del MAb, en un fluido con flujo laminar y perfil parabólico. Para simular la cinética de los Mab en su medio de acción (plasma sanguíneo), se resolvieron numéricamente ecuaciones convencionales por medio de cálculos del movimiento de una partícula con simetría esférica en un fluido laminar de perfil parabólico, utilizando para ello el algoritmo de Verlet de dinámica molecular, con el fin de encontrar la evolución temporal de la velocidad del anticuerpo en el plasma sanguíneo en función del radio, la masa y la densidad de los MAb, y de la presión del fluido en los vasos sanguíneos (vena). Se encontró que en el caso donde se fijo el valor de la densidad del Mab, su cinética aumento cuando la caída de presión en los vasos aumentó. Cuando fijamos la presión en los vasos se encontró que al disminuir el radio del MAb su cinética aumenta, además al aumentar la densidad del MAb encontramos que su cinética también aumenta.

**ABSTRACT:** The study of monoclonal antibodies (MAb) is a field of great interest to science medicine, for example, anti-TNF agents (infliximab and adalimumab) represent an important tool for the management of autoimmune and inflammatory disorders. In this work we focus on the physical description of the transport kinetics of MAb in a fluid with laminar flow and parabolic profile. To simulate the kinetics of the MAb, standard equations were solved numerically (using The Verlet algorithm) to calculate the motion of a particle with a spherically symmetric inside of parabolic laminar flow, in order to find the time evolution of the antibody velocity in blood plasma in function of the increase of the radius, mass and density of the MAb, and the fluid pressure in blood vessels. In the case of we fixed the value of the antibody density, their kinetics increased when the pressure in the vessels increased. When we fixed the pressure in the vessels we found: if we reduce the antibody radius their kinetics increased, and when we increase antibody density we found that their kinetics also increased.

## ¿CÓMO SE AFECTA LA CINÉTICA DE ANTICUERPOS MONOCLONALES POR LA VARIACIÓN DE PARÁMETROS FÍSICOS?

### INTRODUCCIÓN

Los anticuerpos monoclonales son proteínas de alto peso molecular dirigidas contra un determinante antigénico único[1]. Su producción resulta de la hibridación de células de mieloma y de linfocitos sensibilizados, proceso iniciado por Kohler y Milstein en 1975[2]. La producción de anticuerpos monoclonales presenta una serie de ventajas como lo son la producción ilimitada de anticuerpos muy específicos, la ausencia de requerimientos de antígenos puros para obtener nuevas cantidades de anticuerpos y la facilidad de conservar los clones de hibridomas. Sin embargo, su cinética es mas lenta que la de los medicamentos tradicionales[3], lo cual conlleva a una menor eficacia en aquellas indicaciones que precisan una acción específica en determinadas áreas corporales como ocurre en procesos cancerosos o infecciosos.

A continuación mostraremos los aspectos más relevantes del uso del anticuerpo monoclonal (Infliximab) que utilizamos para hacer este estudio, haciendo énfasis en la Enfermedad de Crohn. Después de esta sección se encuentra la información sobre dinámica de fluidos que nos sirvió de base para realizar el presente estudio.

#### Consideraciones generales sobre el uso del anticuerpos monoclonales en el tratamiento de enfermedad de Crohn

La terapia biológica es actualmente una herramienta importante en el manejo de las enfermedades crónicas sistémicas de base inmunológica, incluye todos aquellos fármacos dirigidos contra dianas terapéuticas específicas, como las implicadas en los mecanismos inmunopatogénicos que conducen a la inflamación y a la lesión tisular, utilizándose actualmente en la práctica clínica las siguientes moléculas: anticuerpos monoclonales antiTNF-a (infliximab, etanercept, adalimumab), anti-CD20 (rituximab), anti-CD2 (alefacept), anti-CD11 (efalizumab), una proteína de fusión (abatacept) y el antagonista del receptor humano de la IL-1(anakinra). También se están desarrollando otros compuestos biológicos cuyo fin es inhibir el tráfico linfocitario, entre ellos cabe destacar: los anticuerpos monoclonales contra las integrinas α4 (natalizumab) y α47 (LDP-02), así como la secuencia antisense para la molécula de adhesión intercelular ICAM-18 (ISIS 2302, alicaforsen)[4].

El Infliximab es un anticuerpo monoclonal quimérico, 75% humano y 25% murino, de clase IgG1 que neutraliza la actividad biológica del TNFα, al unirse con gran afinidad a las formas solubles y transmembrana de esta citocina e impedir la unión de ésta a sus receptores. El Adalimumab es un anticuerpo monoclonal que se une específicamente al TNF y neutraliza su función biológica al bloquear su interacción con los receptores p55 y p75 del TNF en la superficie celular, también modula la respuesta biológica inducida por el TNF, produciendo cambios en los niveles de las moléculas de adhesión responsables de la migración leucocitaria (ELAM-1,VCAM-1, e ICAM-1)[4].

El anticuerpo monoclonal anti- TNF, infliximab, está en el mercado americano desde 1998 y en el mercado europeo desde 1999. Las indicaciones de uso del infliximab de la Agencia Europea del Medicamento (EMEA) del 2007, incluye: la Enfermedad de Crohn, Colitis ulcerosa, Artritis Reumatoide, Espondilitis Anquilosante y Psoriasis. El Adalimumab se utiliza en la artritis reumatoide, la espondilitis anquilosante y la artritis psoriásica desde el año 2003. En el año 2007 fue aprobado del comité científico de la EMEA para el tratamiento de la enfermedad de Crohn severa. Además, el Adalimumab se convierte en el primer medicamento biológico autoadministrado por vía subcutánea para el tratamiento de la enfermedad de Crohn, mientras que el infliximab se administra por en infusión intravenosa[5,6].

La enfermedad de Crohn (EC) es una enfermedad inflamatoria crónica del intestino con un curso de recaída (fase activa) y remisión. Las fases de remisión se caracterizan por la ausencia de síntomas, mientras que en las fases de actividad ó recaída, los síntomas se presentan mayoritariamente. Una vez lograda la remisión, el objetivo principal del tratamiento de la EC es el mantenimiento de esta remisión, para ello, el avance más reciente ha sido en el campo de la terapia biológica. Estudios recientes han demostrado que el infliximab y adalimumab son efectivos para mantener la remisión en la enfermedad de Crohn. El infliximab es eficaz para inducir la remisión en los pacientes con la enfermedad de Crohn que continúan teniendo la enfermedad activa a pesar del significativo uso de los tratamientos convencionales. El adalimumab puede está indicado para los pacientes que desarrollan una reacción alérgica severa a infliximab o aquellos que inicialmente responden a infliximab, pero posteriormente pierden su respuesta[7,8].

El origen exacto de la EC es desconocido, pero se engloba dentro del grupo de las enfermedades inflamatorias intestinales, del cual también forma parte la colitis ulcerosa (CU). La hipótesis más aceptada sobre la causa de la enfermedad inflamatoria intestinal, es que es causada por una respuesta inmunitaria a los antígenos, excesivamente agresiva en el intestino de personas susceptibles genéticamente[9]. Estudios sobre la patogénesis de la EC, la han relacionado con una desregulación persistente en la producción de citocinas que puede llevar a una condición inflamatoria crónica sistémica. Al igual que en artritis reumatoide, la EC se ha descrito asociada a la sobreexpresión de citoquinas especialmente del TNF-α, y también a una baja expresión de citocinas, tales como la IL-10, por lo que ambas, son enfermedades inflamatorias crónicas sistémicas de base inmunológica[4].

La Enfermedad de Crohn (EC), se caracteriza por la inflamación transmural, crónica y recurrente de las paredes del tubo digestivo, casi siempre en el íleon terminal, colon e intestino delgado. Los síntomas gastrointestinales más comunes de la EC son: diarrea, dolor abdominal y rectal, hematoquecia, náusea, vómito, dispepsia y fisuras o fístulas perianales (con mayor frecuencia en pacientes cuyo intestino grueso está afectado). Los síntomas generales son: fiebre, fatiga, pérdida del apetito, pérdida de peso, alteración en el crecimiento de los niños y pocas veces ictericia[10]. La EC y la CU, están asociadas con manifestaciones extraintestinales en aproximadamente el 40% de los pacientes. Las manifestaciones extraintestinales de la EC son principalmente: la artritis o artralgias, conjuntivitis, uveítis y episcleritis; otras menos frecuentes que las anteriores: pioedermia gangrenoso (ulcera crónica ideopática), eritema nodoso, síndrome de Sweet (dermatosis neutrofílica febril aguda), estomatitis aftosa; y muy poco frecuentes: colangitis esclerosante. Las manifestaciones extraintestinales de EC pueden seguir un curso que es paralelo a la actividad de la Enfermedad inflamatoria intestinal o pueden presentarse por separado. Las manifestaciones extraintestinales que se presentan en paralelo con la enfermedad inflamatoria intestinal (por ejemplo, la artritis periférica, la uveítis, el pioderma gangrenoso, eritema nodoso y epiescleritis) generalmente responden a infliximab, en este caso, el régimen de dosificación de infliximab debe ser: dosis de inducción con 5 mg / kg a las 0, 2 y 6 semanas, y luego cada 8 semanas. El infliximab, también fue aprobado por Food and Drug Administration de EE.UU. para el tratamiento de la Espondilitis Anquilosante[11].

**Consideraciones básicas sobre dinámica de fluidos**

El estudio del movimiento de fluidos constituye lo que se denomina dinámica de fluidos. Debido a que los fenómenos considerados en la dinámica de fluidos son macroscópicos, un fluido se considera como un medio continuo. Esto significa que siempre se supone que cualquier elemento de volumen pequeño del fluido es suficientemente grande para contener un número elevado de moléculas. La descripción matemática del estado para todo tiempo de un fluido móvil se efectúa conociendo la velocidad del fluido **v**, y dos magnitudes termodinámicas cualesquiera que pertenezcan al fluido, por ejemplo, la presión P y la densidad ρ. Las fuerzas actuando sobre un elemento de volumen del fluido son,[12]

$$\vec{F} = -\vec{\nabla} P - \rho \vec{\nabla} \varphi + f_{Vis\cos a} \qquad (1)$$

Donde $-\rho \nabla \varphi$ representa las fuerzas externas y sí ρ es constante en todo el fluido se dice que este es incompresible. Así la ecuación de movimiento del fluido viscoso incompresible es de la siguiente forma,

$$-\frac{1}{\rho}\vec{\nabla} P + \eta \vec{\nabla}^2 \vec{v} + \vec{g} = \left(\vec{v} \cdot \vec{\nabla}\right)\vec{v} \qquad (2)$$

Donde se tomo $\vec{\nabla}\varphi = -\vec{g}$, es decir el fluido está en un campo gravitacional y $\eta = \frac{\nu}{\rho}$ es la viscosidad especifica. La relación (2) se conoce como la ecuación de Navier Stokes.

De otra parte de la dinámica de fluidos sabemos que las fuerzas de fricción introducen rotación entre las partículas en movimiento, pero simultáneamente la viscosidad trata de impedir la rotación. Dependiendo del valor relativo de estas fuerzas se pueden producir diferentes estados del flujo. Cuando el gradiente de velocidad es bajo, la fuerza de inercia es mayor que la de fricción, las partículas se desplazan pero no rotan, o lo hacen pero con muy poca energía, el resultado final es un movimiento en el cual las partículas siguen trayectorias definidas, y todas las partículas que pasan por un punto en el campo del flujo siguen la misma trayectoria. Este tipo de flujo fue identificado por O. Reynolds y se denomina laminar, es decir las partículas se desplazan con una distribución de velocidades en forma de capas.

Al aumentar el gradiente de velocidad se incrementa la fricción entre partículas vecinas al fluido, y estas adquieren una energía de rotación apreciable, la viscosidad pierde su efecto, y debido a la rotación las partículas cambian de trayectoria. Al pasar de unas trayectorias a otras las partículas chocan entre sí y cambian de rumbo en forma errática. Este tipo de flujo se denomina turbulento. Reynolds estableció la siguiente relación mediante la cual es posible determinar si un flujo es turbulento o no (es decir laminar) en un tubo de sección transversal circular.

$$N_{Reynolds} = \frac{\rho v d}{\eta} \quad (3)$$

Donde *v* es la velocidad media del fluido y *d* es el diámetro del tubo. Para un número de Reynolds menores que 2000 el flujo es laminar, para mayores que 2000 aumenta la probabilidad de que el flujo sea turbulento[13].

En un fluido laminar la distribución de velocidades puede tener múltiples formas, una de ellas se tiene cuando se corta el flujo en planos paralelos y las partículas que están en el mismo plano tienen la misma velocidad, este perfil se denomina plano. Otra posibilidad es que el lugar sobre el cual las partículas poseen la misma velocidad tenga una forma de parábola, cuando es así se denomina perfil parabólico y con la velocidad de las partículas del fluido ser máxima en el centro y tender´ a disminuir sobre las paredes del tubo (Fig. 1). La ecuación para un perfil parabólico de velocidades es la que mas se aproxima a la forma real del flujo sanguíneo en las venas[13],

$$v = \frac{\Delta P}{4\eta l}\left(R^2 - r^2\right) \quad (4)$$

Donde R es el radio de la vena, r es la distancia de cualquier punto al centro del área transversal de la vena, $\Delta P$ su gradiente de presión y *l* su longitud.

## MATERIALES Y MÉTODOS

### Modelo

Como primera aproximación se considera al anticuerpo monoclonal como una esfera rígida que se mueve al interior de un fluido viscoso (el plasma), con flujo laminar de perfil plano, posteriormente se estudia el caso en el cual el perfil de velocidades es parabólico, que es la distribución mas real de velocidades para el flujo de sangre en las venas. En vista de que el medio es mayormente compuesto por agua (número de Reynolds es 1 y el flujo efectivamente puede ser considerado laminar) y de que el tamaño medio de los anticuerpos y moléculas de TNF es mucho menor que el de los glóbulos, estos últimos se ignoran escogiendo un dominio lo suficientemente pequeño y puede considerarse el fluido, en buena aproximación, como un fluido incompresible, isotrópico y newtoniano en un tubo de paredes rígidas y no permeables (no hay intercambio de masa por las paredes) y describirse mediante la ecuación (2). En la situación real hay otros agentes hidrodinámicos fuera de la fuerza viscosa que pueden ser de interés para describir la dinámica del MAb como lo son los términos de Lift y de difusión. A continuación se revisará la importancia de estos términos sobre la dinámica del MAb. Para los estudios numéricos se consideró que el MAb es inyectado vía intravenosa.

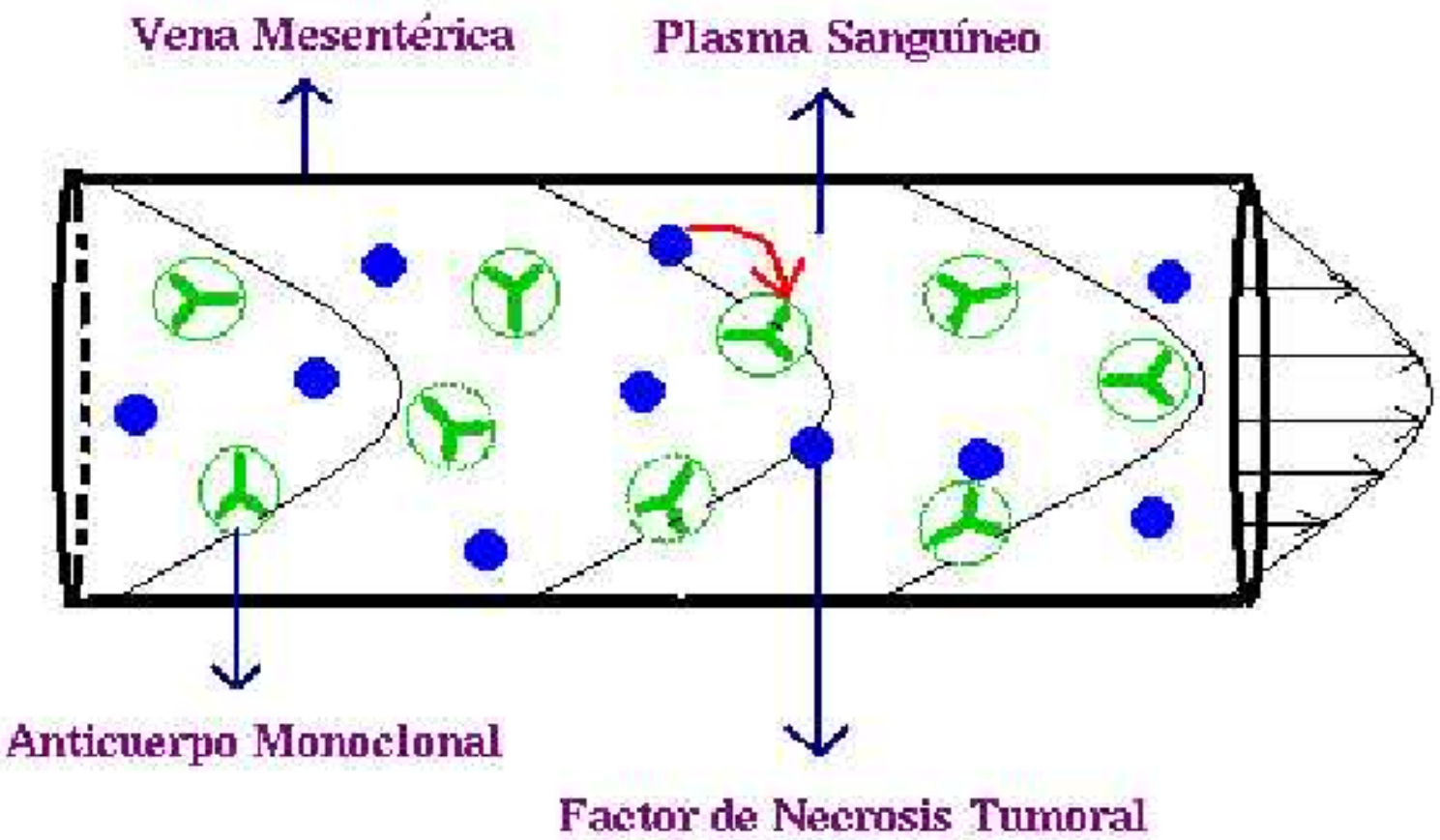

**Figura 1.** Dibujo esquemático que muestra el escenario de estudio y el perfil parabólico de velocidades del flujo.

### Calculo de los términos de Lift y de Difusión presentes en la dinámica del MAb

La fuerza de Lift ocasiona la migración lateral de partículas suspendidas en un fluido, es decir actúa sobre las partículas que se desplazan paralelamente en la vecindad de una pared de un tubo provocando que las partículas esféricas de mayor densidad que el medio viajen hacia las paredes del tubo, mientras que las de menor densidad viajen hacia el centro. Esto explica porque la sangre, siendo una suspensión concentrada, fluye tan fácilmente[14]. La fuerza de lift es función del tamaño del cuerpo, de su distancia a la pared $\delta$ y de la velocidad media del flujo y se describe mediante la siguiente relación[15] ,
$F_{Lift} = d^4 v^2 f(\delta)/\omega^2$ , donde $f(\delta)$ es una función decreciente, complicada, de la distancia $\delta$ a la pared de la tubería (en este caso de la vena), y $\omega$ es una constante de normalización. Examinando esta expresión para unas condiciones fijas de v y de $\delta$, puede verse que esta fuerza es relevante para cuerpos grandes , es decir del tamaño del orden de células como los glóbulos rojos (7.5 micras), mientras que para partículas pequeñas tales como el Infliximab (d = 120 Å ) el termino de Lift le causa un movimiento a lo más de 0.5 veces su diámetro por segundo, es decir $6.47x10^{-9}$m/s.

Por otro lado se sabe que bajo la influencia del movimiento molecular en el fluido, las partículas se mueven de un modo irregular manteniéndolas en suspensión, la distancia media que se mueven en este estado es dada por el coeficiente de difusión. La distancia media que se mueve una partícula por difusión durante un instante es dada por la expresión

$$\bar{r} = 6Dt^{1/2}, D = \frac{K_B T}{3\pi\eta d} \quad (5)$$

Utilizando la anterior relación encontramos que la distancia recorrida en un segundo por el Infliximab teniendo en cuenta el termino de difusión es de 7.76 μm. Así se puede decir que el termino de difusión le causa un movimiento de 647 veces su diámetro por segundo, por esta razón se puede considerar la partícula como browniana y es correcto estudiarla mediante el método de simulación browniana. Además vemos como el término de lift se hace casi despreciable con respecto al de difusión.

**Dinámica Molecular: Algoritmo de Verlet**

El método más ampliamente usado en los programas de dinámica molecular es el *algoritmo de Verlet*. Este usa las posiciones y aceleraciones al tiempo t, y las posiciones del paso previo, *r*(*t-δt*) para calcular las nuevas posiciones en *t+δt*, *r*(*t+δt*). Las relaciones entre posiciones y velocidades en aquellos dos momentos en tiempo pueden ser escritas como[16],

$$\begin{aligned} r(t+\delta t) &= r(t) + \delta t v(t) + \frac{1}{2}\delta t^2 a(t) + \ldots.. \\ r(t-\delta t) &= r(t) - \delta t v(t) + \frac{1}{2}\delta t^2 a(t) - \ldots.. \end{aligned} \quad (6)$$

Estas dos relaciones se pueden sumar para dar,

$$r(t+\delta t) = 2r(t) - r(t-\delta t) + \delta t^2 a(t) \quad (7)$$

Las velocidades no aparecen explícitamente en el algoritmo de Verlet. Estas se pueden calcular de varias formas. Una aproximación muy simple es el dividir la diferencia en las posiciones en tiempos *t+δt* y *t-δt* by 2*δt*, es decir,

$$v(t) = [r(t+\delta t) - r(t-\delta t)]/2\delta t \quad (8)$$

**RESULTADOS Y DISCUSION**

Los resultados que se muestran en esta sección fueron obtenidos teniendo en cuenta los siguientes valores: δt = 0.0001 ps, d = 60 A (Radio del MAb en Armstrong), R = 1.5 $e^7$ Å (Radio de la vena en Armstrong), D = 3.01 $e^{-3}$ $Å^2$/ps(Constante de Difusión), 1.5$e^5$ uma (Masa del MAb), η = 75.6 uma/ $Å^2$ ps, ΔP = 6.08$e^{-11}$ uma/ Å $ps^2$ (Caída de presión en la vena mesentérica). Así se obtuvo valores de velocidad del MAb en Armstrong/ps.

Primero se procedió a calcular la velocidad del MAb como función del tiempo en un flujo laminar plano sin fuerzas aleatorias con el objeto de reproducir los resultados encontrados en la Ref. 3. El resultado obtenido en este caso se muestra en la figura 2.

Observando la figura 2 nos damos cuenta que reproduce muy bien los resultados mostrados en la Ref. 3., lo que nos proporciona seguridad sobre el algoritmo utilizado para este tipo de dinámicas. A diferencia de los resultados mostrados en la Ref. 3 (correspondientes a la velocidad relativa del MAb con respecto al fluido) en este caso podemos ver que la velocidad del MAb tiende a la velocidad del fluido (la cual se tomo de forma arbitraria como 15 cm/s).

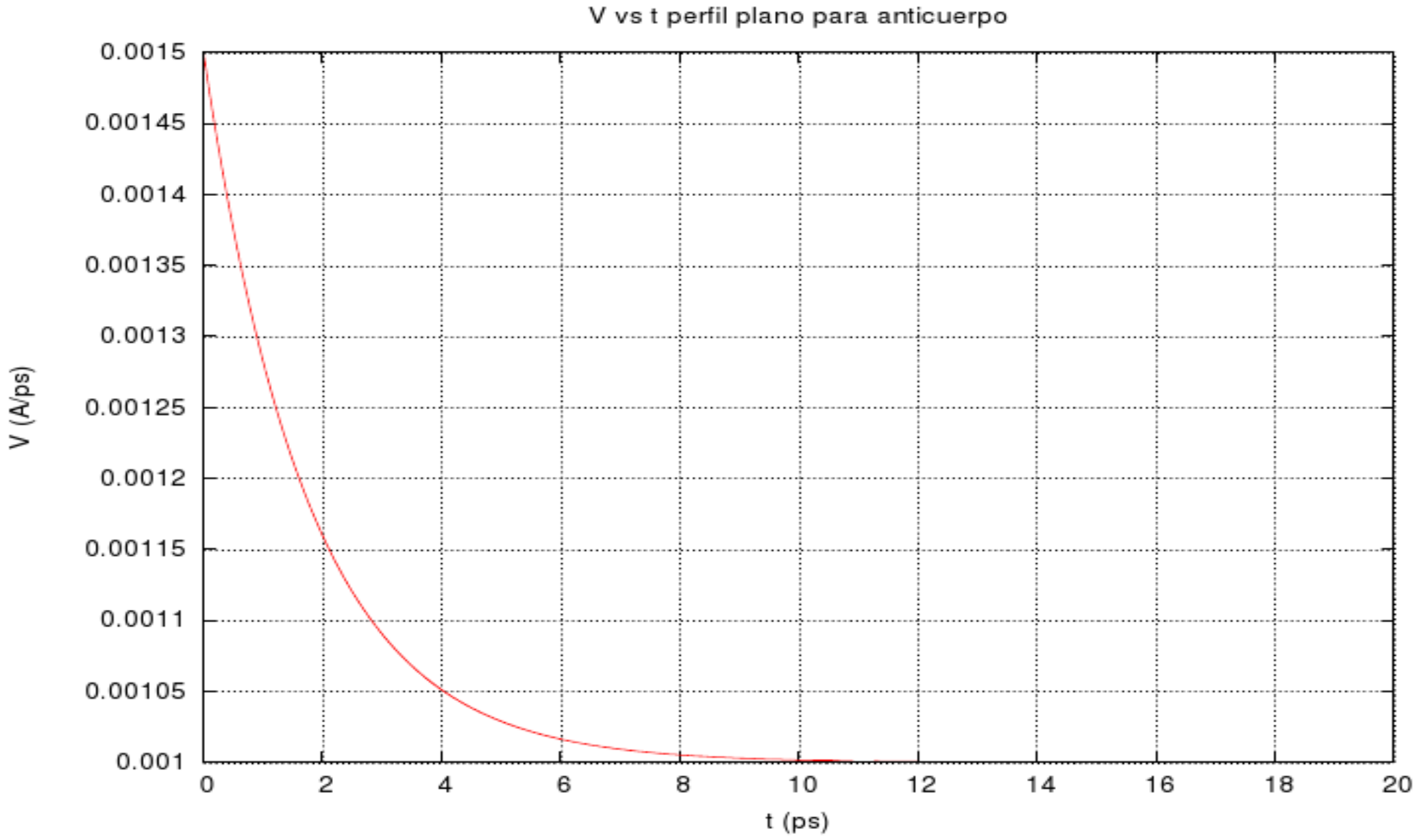

**Figura 2.** Velocidad del MAb como función del tiempo en un fluido de perfil plano.

Debido a que el caso del perfil plano no corresponde exactamente a las condiciones reales del flujo sanguíneo, entonces el segundo paso del presente trabajo fue estudiar la dinámica del MAb en flujo de perfil parabólico implementando para ello de nuevo el algoritmo de Verlet. En la figura 3 se muestra el resultado obtenido. Comparando las figuras 2 y 3, nos damos cuenta que el disminución de la velocidad del MAb en un perfil parabólico es menor que en un perfil plano. Además a partir de estas graficas podemos ver que un flujo con perfil parabólico favorece más la dinámica del MAb.

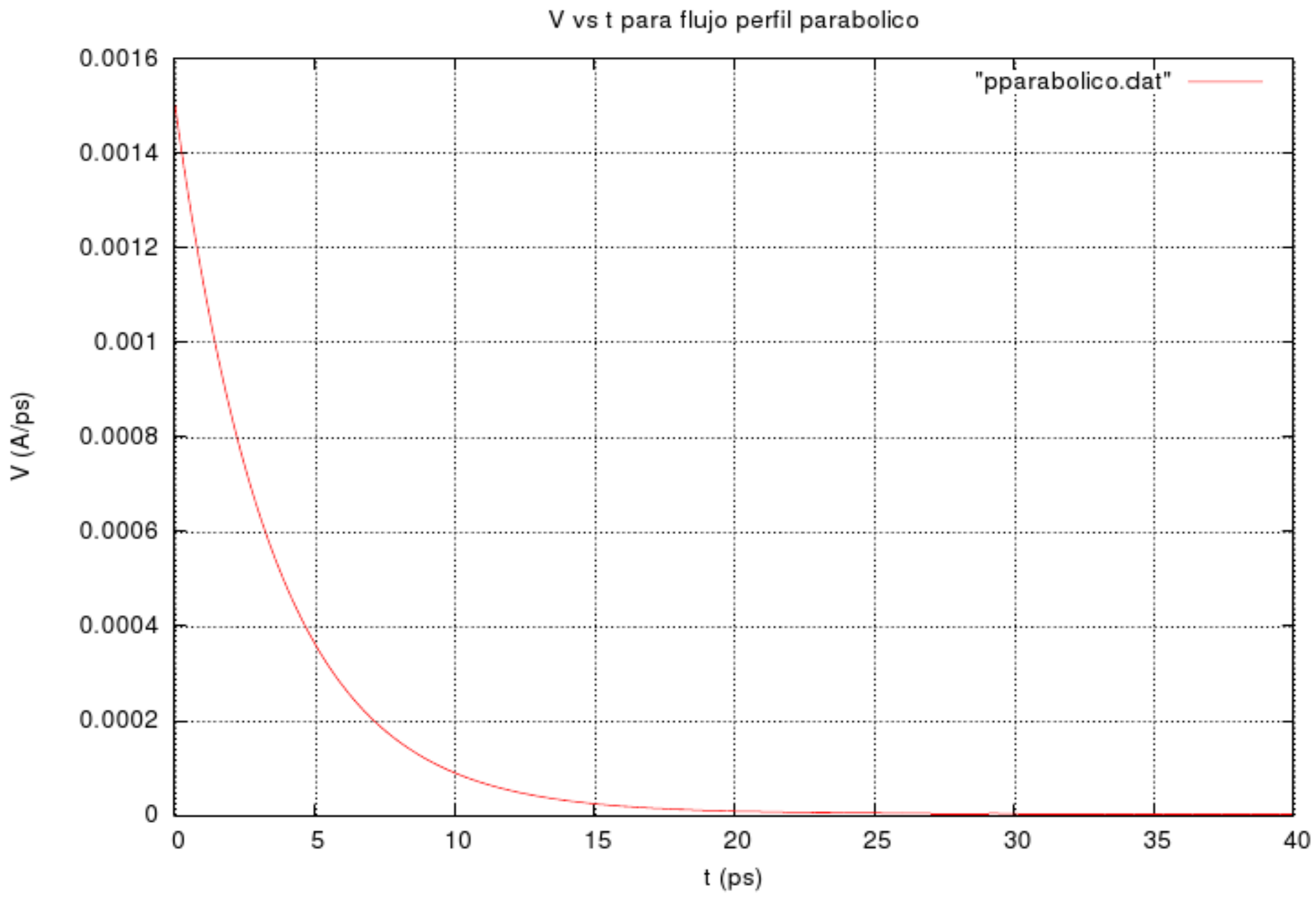

**Figura 3.** Velocidad del MAb como función del tiempo en un fluido de perfil Parabólico.

**Dinámica del MAb (Infliximab) en un fluido laminar de perfil parabólico para diferentes valores de sus parámetros físicos**

La motivación principal de esta investigación es poder determinar cualitativa y cuantitativamente la relación que existe entre la variación de los parámetros físicos de los MAb y el mejoramiento de su cinética. Para ello se estudio la dinámica del MAb para diferentes valores de su radio, de su densidad y para diferentes valores de la caida de presión en la vena donde se mueve el MAb.

En primera instancia se estudio la dinámica del MAb para cuatro valores distintos de caída de presión en la vena, dejando fijo el radio y la densidad del MAb. Estos valores fueron: Presión 1= $6.08e^{-10}$ *uma/ Å*, Presión 2 = $6.08e^{-12}$ *uma/ Å*, Presión 3 = $6.08^{e-11}$ *uma/ Å*, y Presión 4= $6.08e^{-13}$ *uma/ Å.*

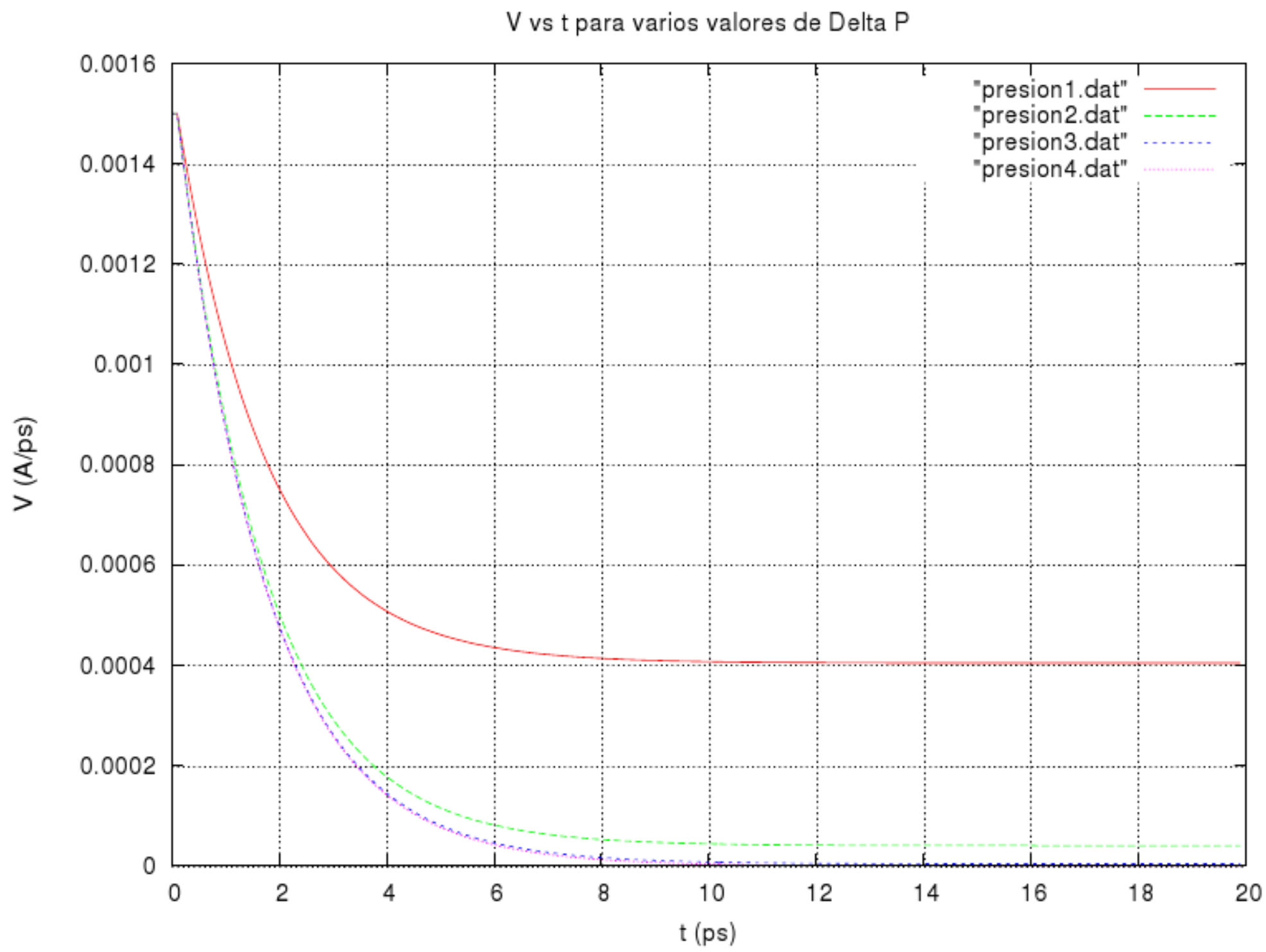

**Figura 4.** Dinámica del MAb en función de la variación de la caída de presión en la vena.

Observando la grafica nos damos cuenta que a mayor presión (Presión 1 = $6.08e^{-10}$ uma/ Å) la cinética del MAb aumenta, es decir que para el mismo intervalo de tiempo el cambio de la velocidad es más pequeño. También se observa que para diferentes valores de caída de presión en la vena, la velocidad del MAb tiende a distintos límites. Es decir que según sea el valor de la caída de presión, cada curva tiene a la velocidad máxima asociada a su perfil parabólico. Cada valor límite es distinto y, está acorde con el resultado mostrado en la ecuación (4), validando así de nuevo nuestros resultados.

En la figura 5 se muestra como cambio la dinámica del MAb al variar el radio del mismo. Los valores escogidos fueron: radio 1 = 70 Å, radio 2 = 60 Å, radio 3 = 50 Å. En esta grafica (figura 5) se observa que al disminuir el radio del MAb se incrementa su cinética, lo cual se esperaba puesto que un cuerpo pequeño produce menos resistencia a su movimiento en el fluido que un cuerpo con mas área de contacto frontal con el fluido. En esta gráfica también se puede ver que las curvas tienden todas a un mismo límite. Lo cual es explicable dado que de nuevo por la ecuación (4) nos damos cuenta que la velocidad la partícula en el fluido parabólico no depende de su radio.

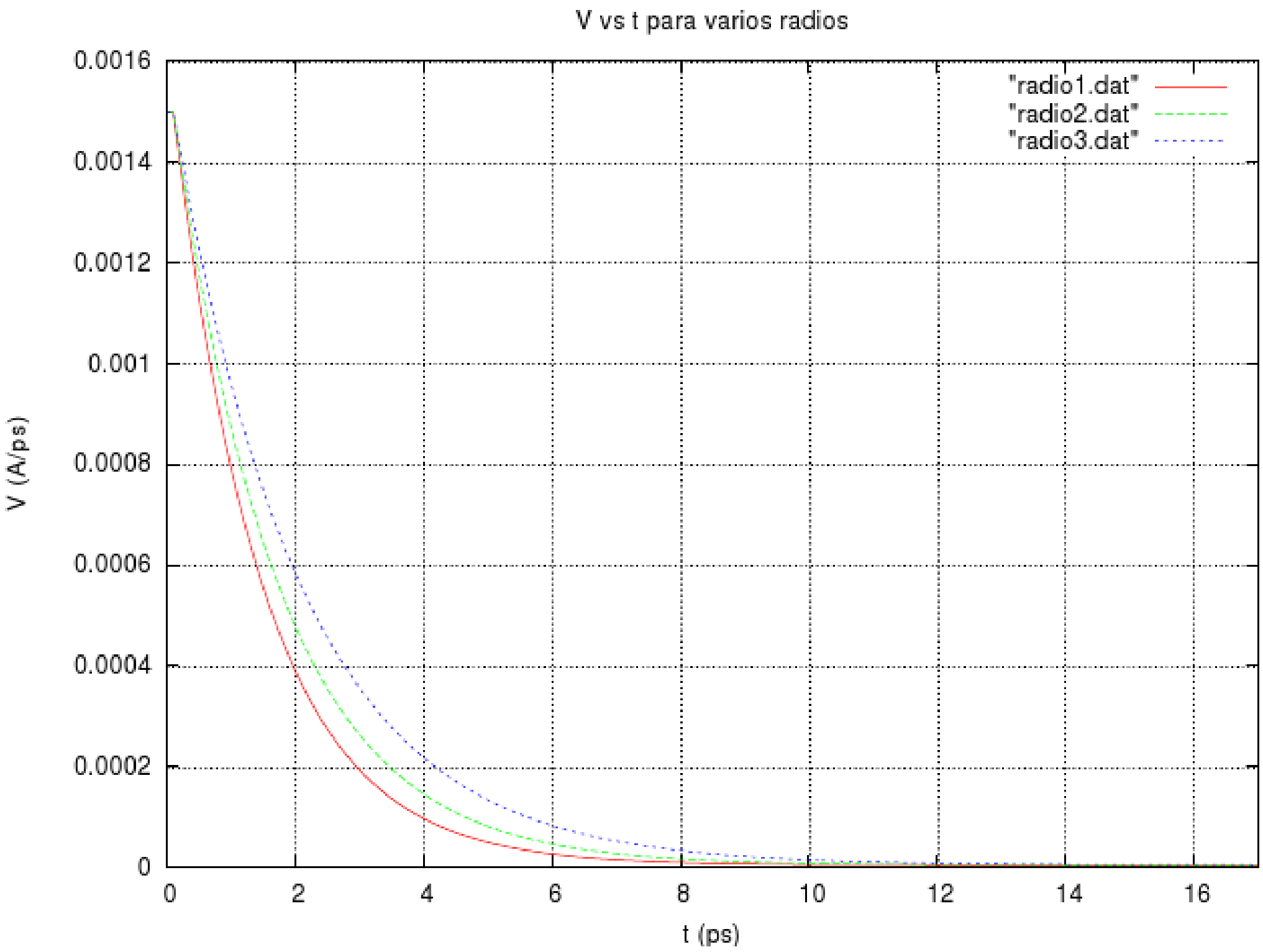

**Figura 5**. Dinámica del MAb en función de la variación de su radio.

Finalmente implementado de nuevo el algoritmo de Verlet se realizó el estudio de la dinámica del MAb para el caso donde se varió los valores de densidad del MAb. Donde $\rho_1$ = 0.142 uma/ $Å^3$ , $\rho_2$ = *0.165 uma/* $Å^3$*,* $\rho_3$ = *0.209 uma/* $Å^3$*.* Analizando esta grafica (figura 6), nos damos cuenta que la cinética de transporte del MAb se incremento a medida que su densidad aumento. Siendo nuestra partícula una esfera, este resultado está muy de acuerdo con el resultado obtenido en la figura 5.

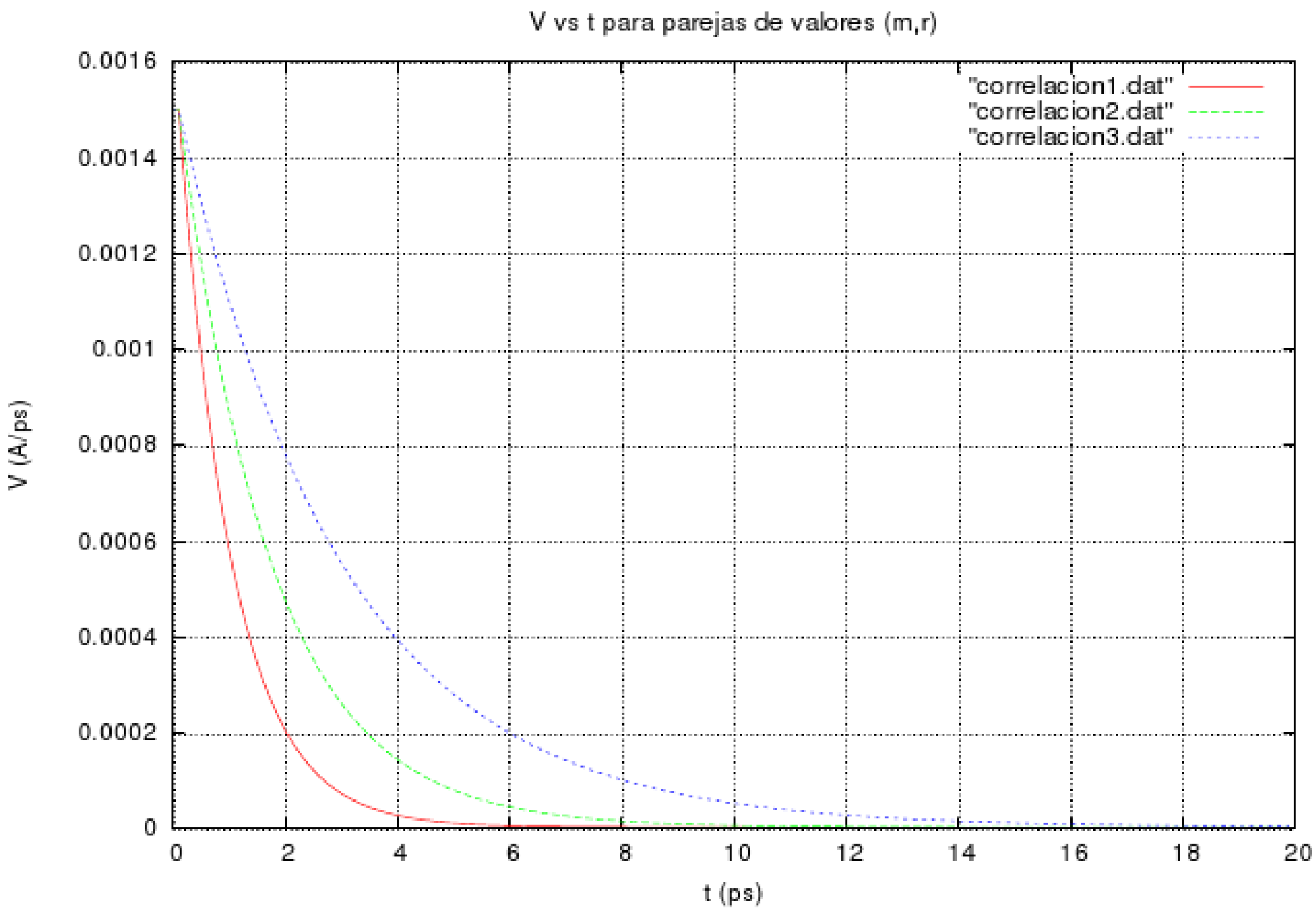

**Figura 6.** Dinámica del MAb en función de la variación de su densidad.

## Anzat para la simulación de varios MAbs

El anzat propuesto por nosotros, para simular la dinámica del MAb en presencia de los mayores componentes del plasma sanguíneo, además de incluir al antígeno correspondiente, consiste en utilizar las distancias y los potenciales de interacción Ag-Ac (Antígeno-Anticuerpo),[17,18,19,20] como parámetros para definir los pesos estadísticos asociados a la interacción entre el Ac (Anticuerpo) y sus vecinos cercanos (i.e. demás partículas del plasma), para luego poder definir las distancias mínimas a las cuáles se presenta interacción entre los anticuerpos y los posibles blancos. La escogencia de los posibles valores de los pesos estadísticos, estarán asociados a cada interacción del anticuerpo con los mayores componentes del plasma (glóbulos rojos, plaquetas) y con los posibles antígenos, esta relación se definirá de manera proporcional a la intensidad de la energía de interacción entre ellos; en otras palabras, significa que el mayor peso estadístico se le va a asignar a la partícula que tenga más probabilidad de unirse con el anticuerpo.

Con este método podemos incluir la información de la especificidad del MAb para reconocer su antígeno (es decir parte de la información de su cinética química), y una información adicional sobre la dinámica de interacción del Macrófago con el MAb, célula encargada en el proceso inmunológico de detectar y digerir al antígeno que se ha acoplado con el anticuerpo. Es decir que el MAb actúa como un sensor biológico de la respuesta inmune, que por delante (Fragmentos FAB) detecta al antígeno y por detrás (FC) es identificado por el macrófago.

De esta manera, nuestro modelo de transporte, además de simular el movimiento de varios anticuerpos en el plasma sanguíneo con la presencia de glóbulos, plaquetas, antígenos y macrófagos, podrá incluir propiedades farmacológicas de los MAb que permitan predecir propiedades físicas y químicas del mismo que contribuya a que los MAb actúen más rápido luego de ser administrados y presenten un tiempo de efecto prolongado.

## CONCLUSIONES

Los MAb han resultado ser efectivos en el manejo de trastornos autoinmunes e inflamatorios tales como la Enfermedad de Crohn, según muestran estudios recientes sobre su efectividad y posibles usos. Por tal razón, es necesario conocer más sobre la cinética de los MAb que permitan mejorar aun mas su eficacia, mejorando su aplicabilidad, por lo que es relevante caracterizar los principales parámetros físicos que están involucrados en el comportamiento cinético de los MAb y estudiar la interacción que éstos presentan con el antígeno.

Los autores de este trabajo consideramos que la cinética general de los MAb presenta dos grandes componentes: cinética de transporte y cinética química. Se mostro mediante el uso del algoritmo de Verlet el hecho de que la cinética de transporte del MAb puede mejorar con una mayor caída de presión en la vena (lo cual es consecuente con el principio de Bernoulli), y con un menor radio y densidad asociados del MAb. Por otro lado, para dar predicciones más acertadas de la dinámica general del MAb que nos lleven a producir un mejoramiento de la efectividad del tratamiento con el medicamento está en complementar nuestro modelo con un estudio acerca de cómo mejorar la velocidad de la reacción Ag-Ac (Antigeno-Anticuerpo). De otro lado, al unir la teoría de interacciones entre proteínas con este trabajo, se puede realizar una simulación de la cinética de varios anticuerpos monoclonales en el torrente sanguíneo siempre y cuando se cuente con los datos experimentales adecuados. Finalmente de esta investigación surgen los siguientes interrogantes: ¿Se seguirán manteniendo estas tendencias en la cinética de los MAb en otras geometrías del anticuerpo que se aproximen a su forma geométrica **Y**?, ¿Será posible medir en el laboratorio las predicciones de tipo teórico hechas en este trabajo?

**AGRADECIMIENTOS**

Especialmente al Dr. José Manuel Lozada, Dra. Norma Valencia y al Dr. Julien Wist por sus fructíferas discusiones.

**REFERENCIAS**

1. Bernard H., "*Diagnostico y tratamiento clínico por el laboratorio*". Salvat Editores. Barcelona. 1998.
2. Milstein C., "*Anticuerpos Monoclonales*". Inmunología, Scientific American. 1982.
3. Delgado C., et al., "*Sobre la cinética de los anticuerpos monoclonales*", Revista CENIC Ciencias Biológicas, Vol 36, No. Especial, 2005.
4. Fernández-Cruz E., et al., "*Introducción a los fármacos biológicos*". Actas Dermosifiliogr., 99(Supl4):2-6.ISSN:0001-7310, 2008.
5. Graves J., et al., "*Off-label uses of biologics in dermatology: rituximab, omalizumab, infliximab, etanercept, adalimumab, efalizumab, and alefacept*", J. Am. Acad. Dermatol.,56:e55-79, 2007.
6. Kuek A., et al., "*Immune-mediated inflammatory diseases (IMIDs) and biologic therapy: a medical revolution*", Postgrad. Med. J.,83:251-60, 2007.
7. Akobeng A. "*The evidence base for interventions used to maintain remission in Crohn's disease*", Aliment Pharmacol Ther., Jan 1;27(1):11-8, 2008.
8. Akobeng A. "*Crohn's disease: current treatment options*", Arch Dis Child., Sep;93(9):787-92, 2008.
9. Podolsky D., "*Inflammatory bowel disease*", N Engl J Med., 347:417- 429, 2002.
10. James S., "*Inmunologic, gastroenterologic, and hepatobiliary disorders*", J. Allergy Clin. Inmunol., 111(suppl 2): S645-658, 2003.
11. Siemanowski B., et al., "Efficacy of infliximab for extraintestinal manifestations of inflammatory bowel disease", Curr. Treat. Options. Gastroenterol.,Jun;10(3):178-84, 2007.
12. Landau et al., "*Mecánica de Fluidos*", Reverte. Capitulo 2. 1991.
13. Kane J., et al., "*Física*", Reverte. Capitulo 14. 2000.
14. Hoyos M., "*Hydrodynamic Separation of Macromolecules, Particles and Cells*", Acta Biologica Colombiana, Vol. 8 No. 1, 11. 2003.
15. Vasseur P., et al.,, "T*he Lateral Migration of a Spherical Particle in TwoDimensional Shear Flow"*, J. Fluid Mech. 78. (385-413), 1976.
16. Garduño-Juárez R., "*Dinámica Molecular*", Notas de Clase, clase 11, Instituto de Ciencias Físicas, UNAM, 2009.
17. Perelson A., et al., "*Immunology for Physicists*", Reviews of Modern Physics, Vol. 69, 4, pp. 1219-1267, 1997.
18. Leckband, et al., "*Forces in Biology*", Quarterly Reviews of Biophysics, 34, 105-267, 2001.
19. Amit A., et al., "*Three-dimensional structure of an antigen-antibody complex at 2.8 A resolution*", Science, Vol. 233, 747-753. 1986.
20. Van Holde, "*Principles in physical biochemistry*", Prentice Hall, New York, 1998.